\documentclass[aps,showpacs,graphicx,twocolumn]{revtex4}%,,preprint
\usepackage{amsmath}
\usepackage{amscd}
\usepackage{graphicx}
\usepackage{color}

\begin{document}

\title{High-order harmonic generation and multi-photon ionization of ethylene in laser fields}

\author{Z.P. Wang,$^{1,2,3}$ P.M. Dinh,$^{3}$ P.-G. Reinhard,${^4}$ E. Suraud${^3}$ and
F.S. Zhang$^{1,2}$\footnote{Corresponding author. Email:
fszhang@bnu.edu.cn}}
\address{$^1$The Key Laboratory of Beam Technology and Material
Modification of Ministry of Education, College of Nuclear Science
and Technology, Beijing Normal University,
Beijing 100875, People's Republic of China\\
$^2$Beijing Radiation Center, Beijing 100875, China\\
$^3$Laboratoire Physique Quantique (IRSAMC), Universit$\acute{e}$ P.Sabatier, 118 Route de Narbonne, 31062 Toulouse, cedex, France\\
$^4$Institut f$\ddot{u}$r Theoretische Physik,
Universit$\ddot{a}$t Erlangen, Staudtstrasse 7, D-91058 Erlangen,
Germany}
\date{\today }

\begin{abstract}
Applying time-dependent local density approximation (TDLDA), we
study the high-order harmonic generation (HHG) of ethylene
subjected to the one-color ($\omega=2.72$~eV) and the two-color
($\omega_1=2.72$~eV and $\omega_2=5.44$~eV) ultrashort intense
laser pulses. The HHG spectrum of ethylene in the one-color laser
field shows the obvious plateaus and odd order harmonics are
produced while the two-color laser field can result in the
breaking of the symmetry and generation of the even order
harmonic. The ionization probabilities are obtained showing the
increase of the ionization probability of higher charge state by
the two-color laser field. The temporal structures of HHG spectrum
of ethylene is explored by means of the time-frequency analysis
showing new insights of the HHG mechanisms in the one-color and
the two-color laser fields.
\end{abstract}
 \maketitle

\section{introduction}
The interaction between strong laser and atoms and molecules is a
hot topic. A series of nonlinear phenomena appear when atoms and
molecules are subjected to intense laser pulse
\cite{Robert,K.Yam,R.Bart,J.H.}. High-order harmonic generation
(HHG) is one of the most studied effects among these nonlinear
phenomena with the results that the coherent emitted light can be
harnessed to produce new coherent trains of extremely short pulses
on the time scale of electron motion \cite{P.B,P.Ago}. A simple
model \cite{Corkum}, based on the idea of tunnelling ionization of
an electron in an atom allows for interpreting the harmonic
generation process in terms of recolliding electron trajectories
and leads to a universal cutoff law for the maximum in photon
energy in harmonic generation. The extra degrees of freedom and
nonspherical symmetry of molecules lead to a further new phenomena
such as molecular high-order harmonic generation (MHOHG)
\cite{P.B, P.Ago, J.Itatani}, charge resonance enhanced ionization
(CREI) \cite{T.Zou, T.Seideman} and bond softening \cite{G.Yao,
SK}. The theoretical understanding of these mechanisms needs to
solute the time-dependent Schr\"{o}dinger equation for all
electrons and all nuclear degrees of freedom. However, this kind
of numerical solutions only exist for smallest systems, such as
atoms \cite{JPHansen, Parker}, H$_2$ and H$_2^+$ \cite{Chelkowski,
Kreibich, ADBandrauk, XBB, Liang}. For large systems the
computation is quite expensive due to the number of degrees of
freedom. In contrast, the well tested time dependent Density
Functional Theory at the level of the time-dependent local-density
approximation (TDLDA) \cite{NATO} serves as a powerful tool to
study the electron dynamics of multi-electron systems
\cite{F.Calvayrac,Takashi,Ben-Nun}. Recently, Madsen \emph{et al}
\cite{C.B.Madsen} studied MHOHG of polyatomic molecules using a
quantum-mechanical three-step model.

The application of two-color laser field for HHG is a fascinating
topic of research science. One can control the formation of
harmonic spectrum and yields of ion or molecular fragments by
changing the relative phase between the fundamental frequency and
its second harmonic fields in the two-color laser field. At
present, in experiment and theory, two-color laser field has been
applied to atoms \cite{Kim,Schumacher,Zhang}, molecules
\cite{SSheely,Chelkowski,Haruimiya,Xuebin,Chen} and clusters
\cite{Nguyen,F.S.Zhang,Y.P.}. However, much less has been done for
more complicated molecules. In this paper we apply TDLDA,
augmented with an average-density self-interaction correction
(ADSIC) \cite{C.Legrand} to investigate theoretically the
ionization and MHOHG of ethylene in one-color and two-color laser
fields.

\section{Theory}
In this section, we represent the real-time method in TDLDA. The
molecule is described as a system composed of valence electrons
and ions. The interaction between ions and electrons is described
by means of a norm-conserving pseudopotentials.

Valence electrons are treated by TDLDA, augmented with an
average-density self-interaction correction (ADSIC)
\cite{C.Legrand}. They are represented by single-particle orbitals
$\phi_{j}(\textbf{r},t)$ satisfying the time-dependent Kohn-Sham
(TDKS) equation \cite{10},
\begin{eqnarray}
i\frac{\partial}{\partial{t}}\phi_{j}(\textbf{r},t) & = & \
\hat{H}_{Ks}\phi_{j}(\textbf{r},t) \nonumber\\
& = &
(-\frac{\nabla^{2}}{2}+V_{eff}(\textbf{r},t))\phi_{j}(\textbf{r},t),
\nonumber \\
& & {} j=1,...,N.
\end{eqnarray}
$V_{eff}$ is Kohn-Sham effective potential composed of four parts,
\begin{eqnarray}
V_{eff}(\textbf{r},t)) = V_{ion}(\textbf{r},t)+V_{ext}(\textbf{r},t)\nonumber \\
+V_{H}[n](\textbf{r},t)+V_{xc}[n](\textbf{r},t),
\end{eqnarray}
where $V_{ion} = \sum_{I}V_{ps}(\textbf{r}-\textbf{R}_I)$ is ionic
background potential, $V_{ext}$ is external potential, $V_{H}$
stands for a time-dependent Hartree part and the final part is
exchange-correlation (xc) potential. The electron density is given
by
\begin{eqnarray}
n(\textbf{r},t)) = \sum_{j}|\phi_{j}(\textbf{r},t)|^{2},
\end{eqnarray}
and the Hartree potential $V_{H}[n](\textbf{r},t)$ is defined as
\begin{eqnarray}
V_{H}[n](\textbf{r},t) =
\int{d^{3}r'\frac{n({\textbf{r}}',t)}{|\textbf{r}-{\textbf{r}}'|}}.
\end{eqnarray}
The xc potential $V_{xc}[n](\textbf{r},t)$ is a functional of the
time-dependent density and has to be approximated in practice. The
simplest choice consists in the TDLDA, defined as
\begin{eqnarray}
V_{xc}^{TDLDA}[n](\textbf{r},t) =
d\epsilon_{xc}^{hom}(n)/dn|_{n=n(\textbf{r},t)},
\end{eqnarray}
where $\epsilon_{xc}^{hom}(n)$ is the xc energy density of the
homogeneous electron gas. For $\epsilon_{xc}^{hom}$ we use the
parametrization of Perdew and Zunger \cite{19}. The form of
pseudopotential for covalent molecule is taken from \cite{20}
including nonlocal part.

The ground state wavefunctions are determined by the damped
gradient method \cite{Paris}. The TDLDA equations are solved
numerically by time-splitting technique \cite{22}. For the
nonlocal part contained in Hamiltonian, we deal with it in an
additional propagator and treated it with a third-order Taylor
expansion of the exponential \cite{Paris}. The absorbing boundary
condition is employed to avoid periodic reflecting electrons
\cite{24}.

For the laser field, neglecting the magnetic field component can
be written as
\begin{eqnarray}
E(t)= E_{0}f(t)[cos(\omega(t))+A cos(2\omega(t)+\phi)]
\end{eqnarray}
where $E_{0} \propto {\sqrt{I}}$, $I$ denoting the laser
intensity, $\omega$ is the laser frequency and $f(t)$ is the pulse
profile. $A$ is the electric field strength ration between the two
frequency and $\phi$ is the relative phase. In this paper, $A$ is
chosen 0 and 0.5 respectively for the one-color and the two-color
laser field and $\phi=0$.

The laser induced electron dipole moment can be obtained by
\begin{eqnarray}
D(t)= \int{rn(\textbf{r},t)d^{3}}r
\end{eqnarray}
The Fourier transform of $D(t)$ gives the MHOHG power spectrum
$|D_F(\omega)|^2$.

The harmonics intensity as a function of harmonic frequency
$\omega$ and emission time $\beta$ can be obtained by the
time-frequency analysis \cite{Antoine}
\begin{eqnarray}
D_G(\omega,\beta)=|\int D(t)e^{i\omega
t}e^{-(t-\beta)^2/2\alpha^2}dt|^2
\end{eqnarray}
where the window function width $\alpha$ is chosen as one-tenth of
the laser optical period.

The number of escaped electrons is defined as
\begin{eqnarray}
N_{esc} = N_{t=0} - \int_{V}d^{3}rn(\textbf{r},t)
\end{eqnarray}
where $V$ is a volume surrounding molecule. A detailed link with
experiments is the probabilities $P^k(t)$ of finding the excited
molecules in one of the possible charge states $k$ to which they
can ionize. The formula can be obtained from \cite{24}.

\section{Results and discussion}

The ethylene is the simplest organic $\pi$ system holding $D_{2h}$
symmetry. In our calculation, there are 12 valence electrons. It
is in $x$-$y$ plane with the center of mass at the origin. The
laser polarization is along $x$ direction which is parallel to the
axis of CC double bond. The laser time profile is chosen the Ramp
envelope. The ramp-on and ramp-off time are both 8~fs and the
duration is 30~fs. For the one-color laser field, $I=
10^{14}$~W/cm$^2$ and $\omega= 2.72$~eV and for the two-color
laser field, $I_1=10^{14}$~W/cm$^2$, $I_2=I_{1}/4$ and
${\omega}_1=2.72$~eV, ${\omega}_2=5.44$~eV. It should be noted
that the two-color laser field is non-symmetric, which is the same
as that in \cite{Xuebin}, the negative and positive amplitudes of
the electric field strength are not equal.

\subsection{Ionization properties}

\begin{figure}[!ht]%[tpb]
\begin{center}
\includegraphics[width=6cm,angle=0]{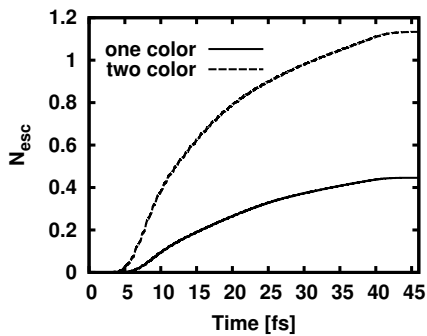}
\caption{Time evolutions of the number of escaped electrons of
ethylene in the one-color and the two-color laser fields. Solid
line: one-color ($\omega= 2.72$~eV) laser field with
$I=10^{14}$~W/cm$^2$. Dashed line: two-color laser field with
${\omega}_1=2.72$~eV, $I_1=10^{14}$~W/cm$^2$ and
${\omega}_2=5.44$~eV, $I_2=I_{1}/4$, the relative phase
$\phi=0$.}\label{escelectron}
\end{center}
\end{figure}

Fig.~\ref{escelectron} shows the time evolutions of the number of
escaped electrons of ethylene in the one-color and the two-color
laser fields. One can find that the electron emission takes the
same pattern in both cases. The ionizations start at around 5~fs
with a constant speed in both cases, while it is obvious that
electrons start to escape a little earlier in the two-color laser
field. Then in both cases electron emissions become more and more
slow. Finally, at around 42~fs they are saturated at 1.18 in the
two-color case and 0.42 in the one-color case, which is 4~fs
earlier than the laser pulses are switched off. Moreover, it is
obvious that the ionization is enhanced by the two-color laser
field.

\begin{figure}[!ht]%[tpb]
\begin{center}
\includegraphics[width=8cm,angle=0]{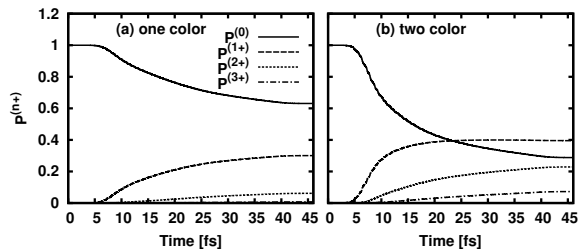}
\caption{The ionization probabilities of ethylene in the one-color
laser field (a) and two-color laser field (b) for the same cases
as in Fig.~\ref{escelectron}.}\label{probab}
\end{center}
\end{figure}

Fig.~\ref{probab} represents ionization probabilities of various
charge states for ethylene in one-color and two-color laser
fields. From Fig.~\ref{probab}(a) we can see that in the first
5~fs, there is mainly neutral ethylene in the one-color laser
field and probabilities of $P^{(1+)}$ and $P^{(2+)}$ are almost
zero. Then from 5~fs to 42~fs, the probability of neutral ethylene
decreases more and more slowly and reaches a saturated value 0.62
before the laser is switched off. During this period, $P^{(1+)}$
and $P^{(2+)}$ increase more and more slowly and they are also
saturated at 42~fs. Their final values are 0.3 and 0.05
respectively. The probability of $P^{(3+)}$ is so small that it is
not visible in the figure. Furthermore, one can find that in the
course of the one-color laser pulse, neutral ethylene predominates
the main probability.

For the ionization probabilities of ethylene in the two-color
laser field, as shown in Fig.~\ref{probab}(b), it is obvious that
the time evolution pattern is different from that in the one-color
case. In the first 5~fs, there is mainly neutral ethylene in the
the two-color laser field. From 5~fs to 42~fs, $P^{(0)}$ decreases
more and more slowly and reaches a saturated value 0.28. This
value is smaller than that in Fig.~\ref{probab}(a). It is
noteworthy that from 5~fs to 10~fs, $P^{(0)}$ drops quickly from 1
to 0.62 in the two-color case while it takes 37~fs for $P^{(0)}$
to drop from 1 to 0.62 in the one-color case. In
Fig.~\ref{probab}(b), $P^{(1+)}$ starts to increase quickly at
around 5~fs and it exceeds $P^{(0)}$ at around 23~fs. Finally, it
is saturated at 0.4. For $P^{(2+)}$, it starts to increase at
around 7~fs, which is a litter later than $P^{(1+)}$, but earlier
than $P^{(3+)}$. The increase trends of $P^{(2+)}$ and $P^{(3+)}$
are quite similar. Both of them increase slowly and reach
saturated values 0.22 and 0.06 respectively at around 42~fs. We
can also find that after 23~fs, $P^{(1+)}$ dominates mainly the
charge state.

From the above discussion we can say that the probability of
higher charge state is larger for the two-color than that for the
one-color laser field. This is related to the fact that the local
maximum of amplitude strength for the two-color laser field is
larger than for the one-color case. This is consistent with the
result from the interaction between atoms and laser fields
\cite{Tong}.

\subsection{Analysis of MHOHG}

\begin{figure}[!ht]%[tpb]
\begin{center}
\includegraphics[width=8cm,angle=0]{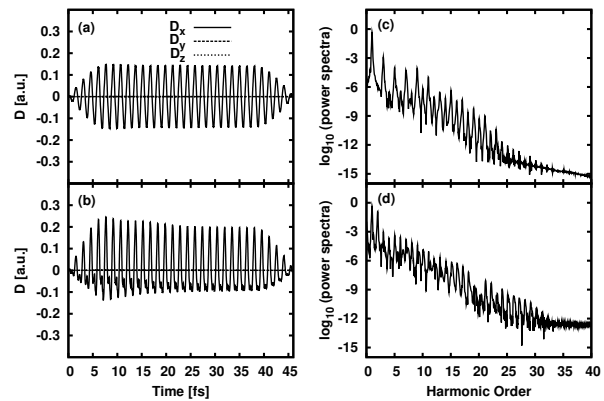}
\caption{(a) and (b): the time evolution of dipole moments of
ethylene along $x$, $y$ and $z$ directions in the one-color and
the two-color laser fields. (c) and (d): MHOHG spectrum of
ethylene in the one-color and two-color laser fields. The laser
parameters are the same as those in
Fig.~\ref{escelectron}.}\label{HHG}
\end{center}
\end{figure}
Figs.~\ref{HHG}(a) and (b) exhibit the time evolution of dipole
moments along $x$, $y$ and $z$ directions of ethylene in the
one-color ($\omega$) and the two-color ($\omega+2\omega$) laser
field respectively. One can find that in both cases the dipole
moments along $x$ direction are distinct while those along $y$ and
$z$ directions are very small. This is due to the laser
polarization is along $x$ direction. However, $D_x$ displays the
different behaviors in two cases. In Fig.~\ref{HHG}(a), $D_x$
exhibits the relative symmetry while in Fig.~\ref{HHG}(b) it is
not symmetric because of the non-symmetric property of the
electric field strength of the two-color laser field.

The MHOHG spectra calculated by our TDLDA in the one-color and the
two-color laser fields are shown in Figs.~\ref{HHG}(c) and
Fig.~\ref{HHG}(d). It is well known that the chosen laser
intensities and frequencies are certain essential physical
parameters which allow for quasistatic interpretations of strong
field laser-atom processes \cite{Corkum, Scrinzi}. In this paper,
ponderomotive energy $U_p=eI/4m\omega^2$ for $\omega=2.72$~eV and
$2\omega=5.44$~eV are 1.93~eV and 0.12~eV respectively. The
Keldysh parameter $\gamma$ separating multiphoton and tunnelling
ionization regimes for ethylene ionization potential $I_p=11.5$~eV
is $\gamma=\sqrt{I_p/2U_p}=1.7$. Thus the present parameters
situate our calculations above the tunnelling ionization regime.

According to the atomic HG cut-off law, $E_{max}=I_p+3.17U_p$, the
cutoff MHOHG of ethylene in the one-color laser field is at 6.
From the MHOHG spectrum in Fig.~\ref{HHG}(c) we can find a clearly
cut-off at the 9th order. Although this value is not exactly the
same as the value predicted by the atomic HG cut-off law, it is
quite near and we attribute the difference to the fact that
ethylene is polyatomic molecule while the cut-off law is for
atoms. One can also find in Fig.~\ref{HHG}(c) that the MHOHG
spectrum of ethylene consists of a series of peaks, first
decreasing in amplitude and then reaching a plateau. The plateaus
is rather short from the 5th order harmonic to the 9th order
followed by continuously decrease. Thus the MHOHG spectrum of
ethylene behaves basically like an atom with recollision of the
electron with the compact molecular ion being the principal HG
mechanism. Furthermore, one can find in Fig.~\ref{HHG}(c) that
ethylene generates odd harmonics.

Compare the harmonic spectrum in Fig.~\ref{HHG}(d) to that in
Fig.~\ref{HHG}(c), we can see that the shapes of MHOHG spectrum
are different. In Fig.~\ref{HHG}(d), the plateaus is not so
obvious and the decrease lasts longer. This is due to the
interference effect of the $1\omega$ and $2\omega$ components of
the two-color laser field. It can also be found that even
harmonics are well produced in the two-color case. One explanation
of this is that even harmonic results from the sum of an odd
number of $2\omega$ photon plus two $1\omega$ photons. The other
interpretation is that even harmonic generation requires broken
reflection symmetry because only this allows that even harmonic
generation transforms a squared dipole excitation into one dipole
signal, i.e., $\hat{D}^2{\longrightarrow}\hat{D}$. That transition
cannot be mediated by a reflection symmetric system because parity
is then conserved, but $\hat{D}$ has negative parity while
$\hat{D}^2$ has positive parity. Ethylene is symmetric and a
two-color laser field breaks the symmetry so that a series of even
harmonics are produced.

\begin{figure}[!ht]%[tpb]
\begin{center}
\includegraphics[width=8cm,angle=0]{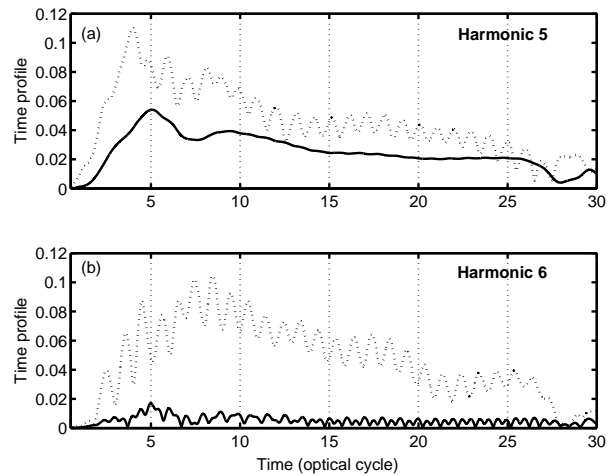}
\caption{Time profiles of the harmonics at about the 5th one and the
6th one in the one-color laser field (solid line) and the two-color
laser field (dotted line). The laser parameters are the same as
those in Fig.~\ref{escelectron}.}\label{Harmonics}
\end{center}
\end{figure}
To investigate the detailed temporal structure of the MHOHG
spectrum of ethylene, we perform the Gabor transform of the dipole
moments in the one-color and the two-color laser field
respectively. The laser parameters are the same as those in
Fig.~\ref{escelectron}. Here we consider two typical harmonics
which are both in the plateau, one is odd harmonic (5th) and the
other one is even harmonic (6th), as shown in
Fig.~\ref{Harmonics}. It should be noted that the optical cycle
refers to the one-color laser frequency. In
Fig.~\ref{Harmonics}(a) we can find that in the one-color laser
field, the time profile of the 5th harmonic shows a relative
smooth function of the driving laser pulse which indicates that
the multiphoton mechanism dominates this lower harmonic regime.
However, for the two-color case, the time profile of the 5th
harmonic is stronger and it exhibits well the one burst within
each optical cycle from the 5th optical cycle to the 25th optical
cycle. This can be attributed to the recollision of the electronic
wave packet with the ionic cores. Due to the first five and the
final five optical cycles are ramp-on and ramp-off times of laser
pulses, there are only two and four bursts in these two parts of
times. For the 6th harmonic, as shown in Fig.~\ref{Harmonics}(b),
it is obvious that in the one-color laser field, the time profile
of the 6th harmonic is very weak, whereas in the two-color laser
field, it is almost two orders of magnitude stronger. Again, it
shows well the one burst within each optical cycle from the 5th
optical cycle to the 25th optical cycle.

\section{Conclusions}
In this paper, we have simulated the MHOHG spectra and ionization
of ethylene induced by the one-color and the two-color laser
fields in the multiphoton regimes with TDLDA. We find that the
ionization and the ionization probability of higher charge state
are enhanced by the two-color laser field. It is shown that the
MHOHG spectrum of ethylene in the one-color laser field exhibits
the typical atom HHG spectrum and odd order harmonics are
produced. The two-color laser field can result in the breaking of
the symmetry and generate the even order harmonics. Furthermore,
the detailed temporal structure of MHOHG spectrum of ethylene is
obtained by means of the time-frequency transform providing new
insights of the MHOHG mechanisms in the one-color and two-color
laser fields.
\section*{ACKNOWLEDGEMENTS}

This work was supported by the National Natural Science Foundation
of China (Grants No. 10575012 and No. 10435020), the National
Basic Research Program of China (Grant No. 2006CB806000), the
Doctoral Station Foundation of Ministry of Education of China
(Grant No. 200800270017), the scholarship program of China
Scholarship Council and the French-German exchange program PROCOPE
Grant No. 04670PG.

\end{document}